\documentclass[10pt,preprint]{aastex}

\bibpunct{(}{)}{;}{a}{}{,}

\slugcomment{Accepted to the ApJL}

\shorttitle{Low Luminosity Outflows}
\shortauthors{Bussmann et al.}

\begin{document}

\title{A CO ($J=3-2$) Outflow Survey of the Elias 29 Region}

\author{R. S. Bussmann, T. W. Wong, A. S. Hedden, C. K. Kulesa, and C. K. Walker}
\affil{Steward Observatory, Department of Astronomy, University of Arizona, 933
N. Cherry Ave., Tucson, AZ 85721}
\email{rsbussmann@as.arizona.edu,twong@email.arizona.edu,ahedden@as.arizona.edu}
\email{ckulesa@as.arizona.edu,cwalker@as.arizona.edu}

\begin{abstract} 

We present a 5$\arcmin \times 5\arcmin$ integrated intensity map of
$^{12}$CO~($J=3-2$) emission from the $\rho$~Ophiuchi cloud core that traces
low-luminosity outflow emission from two protostars: Elias~29 and, most likely,
LFAM~26. The morphology of the outflow from Elias~29 is bipolar and has a
curved axis that traces the S-shaped symmetry seen in H$_2$ 
emission.  The outflow from LFAM~26 is a new detection and oriented in the
east/west direction near the plane of the sky with most of the blue-shifted
emission being absorbed by intervening clouds.  The outflow axis of this object
also appears to intersect a knot of H$_2$ emission previously attributed to
Elias~29.  LFAM~26 is a low luminosity source ($L_{bol} =0.06 \: L_\sun$)
which, in combination with the observed outflow, makes it a candidate Very Low
Luminosity Object (VeLLO).   We derive lower limits to the gas column densities
and energetics for both outflows.  The mechanical luminosities for Elias~29 and
LFAM~26 are 6.4 and 10.3 $\times 10^{-3} \: L_\sun$, respectively.

\end{abstract}



\section{Introduction}
\label{sec:intro}

We present $^{12}$CO~($J=3-2$) observations towards the (36 $L_\sun$) Class~I
protostar Elias~29 ($\alpha = 16^{\rm h}27^{\rm m}09^{\rm s}.5, \;
\delta=-24^\circ37\arcmin18\arcsec$ [J2000.0]) located in the $\rho$~Ophiuchi
star forming region.  The $\rho$-Oph cloud complex is nearby \citep[$d \sim
120\;$pc,][]{2006AAS...20913303L} and has a diverse (proto-)stellar environment
that makes it a superb laboratory for studying the impact of molecular outflows
on the surrounding environment.  The region around Elias~29 is a particularly
well-studied component of the cloud complex.  Two foreground clouds at $v_{LSR}
= 2.7$~km$\:$s$^{-1}$ and $v_{LSR} = 3.8$~km$\:$s$^{-1}$ account for the
presence of cool ice along the line of sight and absorb much of the
blue-shifted outflow emission in this region \citep{2002ApJ...570..708B}.
Furthermore, \citet{2002ApJ...570..708B} identified a ridge of dense material
at $v_{LSR} = 5.0$~km$\:$s$^{-1}$ within which both Elias~29 and LFAM~26
reside.  \citet{2002A&A...395..863C} obtained spectra of the $^{12}$CO~($J=6-5$)
transition with the JCMT and the $J \ge 15$ CO levels with the ISO SWS towards
Elias~29.  These authors found evidence for a super-heated surface disk layer
with a temperature and mass similar to that found in Herbig AeBe stars
\citep{2001ApJ...561.1074T}, suggesting that Elias~29 may be a deeply embedded
Herbig AeBe star or a transition object between the standard Class~I and
Class~II types.  \citet{1996A&A311..858B} reported the discovery of an outflow
around Elias~29 and estimated its momentum rate.  A larger map of the region
was made in the $^{12}$CO~($J=2-1$) transition by \citet{1997ApJ489L..63S}.  

In this paper, we present $^{12}$CO~($J=3-2$) observations that further expand
the size of the mapped region and reaffirm the bipolar nature of the outflow,
but find that it is more extended than previously thought.  Furthermore, the
map has sufficient size to identify a previously undiscovered outflow whose
source appears to be LFAM~26.  First discovered as a radio source
\citep{1991apj...379..683L}, LFAM~26 was later identified as a low luminosity
($L_{bol} = 0.064 \: L_\sun$) Class~I protostar based on ISOCAM photometry
\citep{2001A&A...372..173B}.  The extreme faintness of this source, coupled
with the detection of molecular outflows presented in this paper, make it a
candidate VeLLO.  However, detailed modeling of the infrared SED will be
necessary to determine if this object meets the primary criterion of $L_{int} <
0.1 \: L_\sun$.

\section{Observations}
\label{sec:obs}

We made a 5$\arcmin \times 5\arcmin$ $^{12}$CO~($J=3-2$) On-The-Fly map
centered spectrally at 345.795975 GHz and spatially on Elias~29 using the 10m
Heinrich Hertz Telescope (HHT) on Mt. Graham, Arizona during April 2001 and
January 2006.  A 256 channel filterbank spectrometer with 250~kHz resolution
was used as the backend in 2001, and two 2048 channel Acousto-Optical
Spectrometers (AOS) with 475~kHz resolution were used in 2006.  The
off-position for both maps was located at a clean offset
\citep{2005ApJ625..194K} of $\Delta \alpha = 2.5^\circ, \; \Delta \delta =
-0.83^\circ$ from Elias~29.  We obtained two rows of data per OFF, translating
to $\sim$1 OFF observation per minute for a 10$\arcsec$ per second scan rate.
Each row is offset in declination by 10$\arcsec$.  At 345~GHz, the half-power
telescope beamsize and main beam efficiency were measured to be 22$\arcsec$ and
0.7, respectively.  Pointing accuracy of a few arcseconds was maintained by
observing planets approximately once per hour.  The OTF data grid was convolved
with an 11$\arcsec$ beam to obtain a Nyquist-sampled map.  System temperatures
during the observing run were 500-1500~K.  Line strengths were calibrated using
hot and cold loads in conjunction with sky measurements at the off position
\citep{2001PASP..113..567S}.  Calibration uncertainty is estimated to be $\la
20$\%.  For both datasets, the main beam temperature ($T_{mb} =
T_A^*/\eta_{mb}$) showed a typical rms noise level of 0.7~K .  The 2001 data
are smoothed in velocity space to match the 0.4~km~s$^{-1}$ bins of the 2006
data.  The final datacube used in the analysis is an rms-weighted average of
the two datasets and has a typical rms noise temperature of $T_{mb} = 0.4$~K.

\section{Results}
\label{sec:res}

\subsection{Outflow Morphology}\label{sec:morph}

\begin{figure}
\plotone{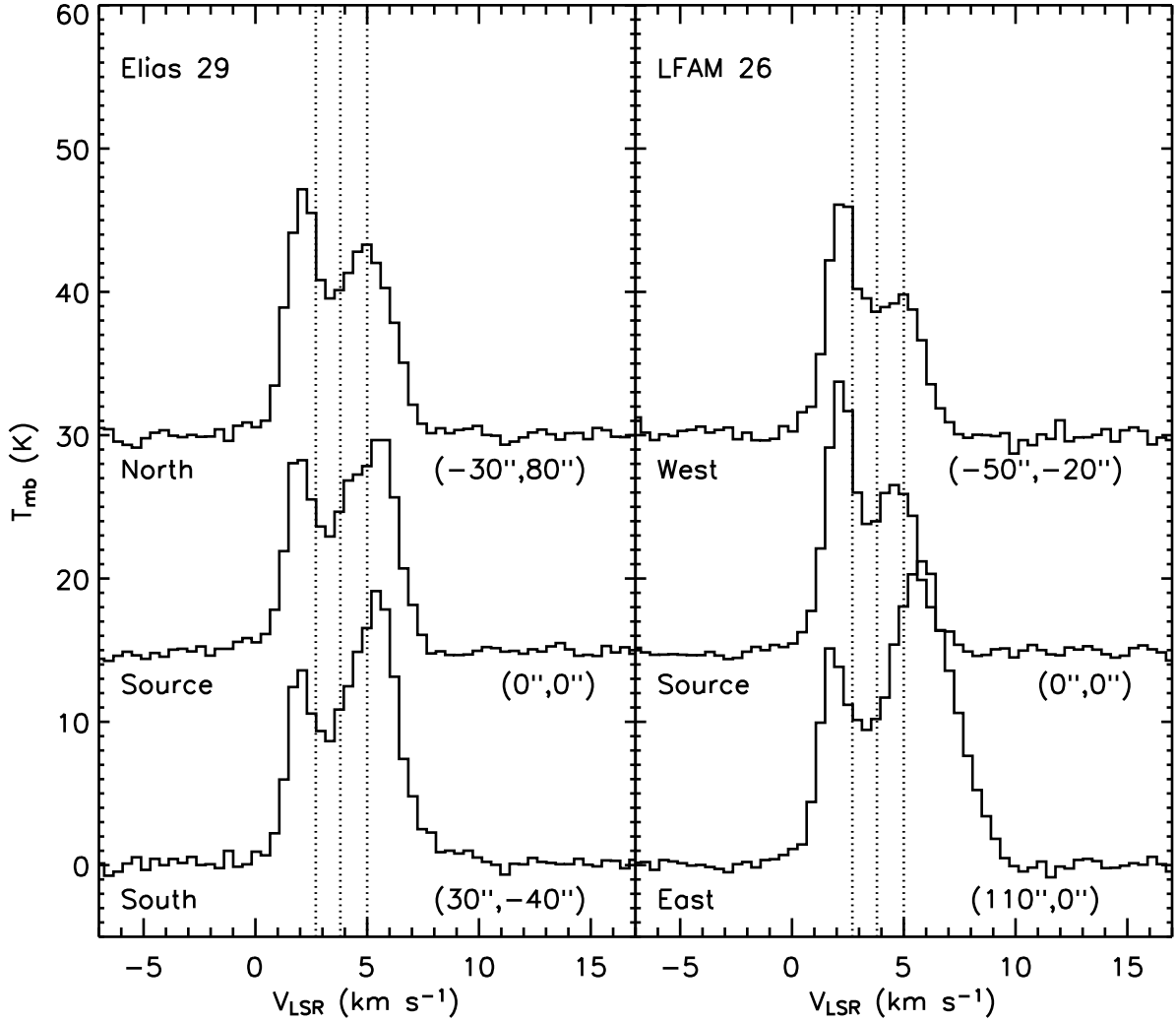}
\caption{Spectra of Elias~29 and LFAM~26 in addition to their respective
outflows.  The spatial offsets from the driving sources are given below each
spectrum (north and east are positive).  LFAM~26 is offset from Elias~29 by
(-60$\arcsec$, 44$\arcsec$).  Dotted lines indicate the velocity of intervening
absorption clouds at $v_{LSR} = 2.7$ and $v_{LSR} = 3.8$~${\rm km \: s}^{-1}$
as well as an absorption ridge at $v_{LSR} = 5.0$~${\rm km \: s}^{-1}$ detected
by \citet{2002ApJ...570..708B}.  The spectra have been shifted in intensity by
15~K.} \label{fig:spectra}
\end{figure}

In Figure~\ref{fig:spectra}, we show spectra located at the driving sources as
well as at the approximate peak of the integrated emission intensity in the
blue and red lobes of both outflows.  The vertical lines demarcate the position
in velocity space of intervening absorption clouds along the line of sight at
$v_{LSR} = 2.7$~km$\:$s$^{-1}$ and $v_{LSR} = 3.8$~km$\:$s$^{-1}$, as well as a
ridge of material at $v_{LSR} = 5$~km$\:$s$^{-1}$ \citep{2002ApJ...570..708B}.
Each of the spectra show signs of significant absorption due to these
intervening clouds.  

\begin{figure}
\epsscale{0.90}
\plotone{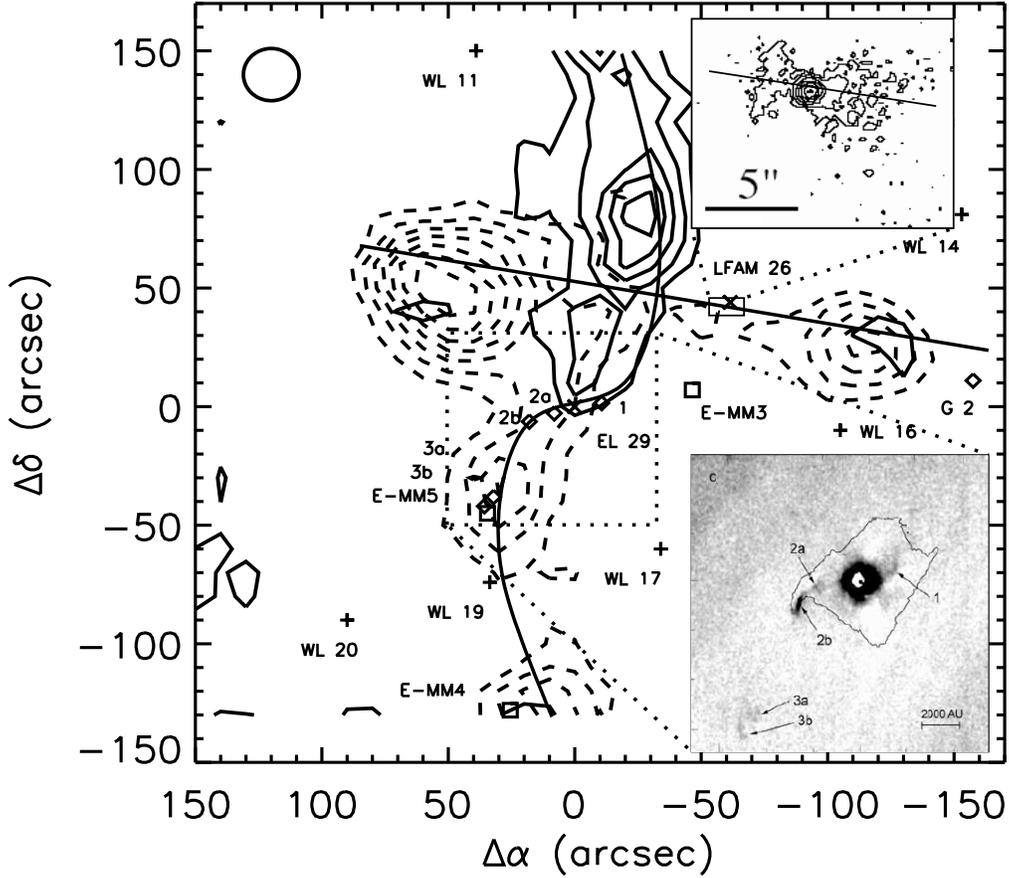}
\caption{Integrated intensity map of the blue-shifted (solid contours, $v_{LSR}
= -4$ to 1 km~s${^-1}$) and the red-shifted (dashed contours, $v_{LSR} = 7$ to
12 km$\:$s$^{-1}$) outflow emission centered on Elias~29.  Lowest contour shown
represents $3\sigma$, steps are in units of $2\sigma$.  $\sigma_{bw} =
\sigma_{rw} = 1.4 \:$K$\,$km$\:$s$^{-1}$.  Crosses show the location of
Elias~29 and LFAM~26, plus signs indicate the position of infrared sources
\citep{1983ApJ274..698W}, squares represent starless cores
\citep{1998A&A...336..150M}, and diamonds show H$_2$ emission knots from
\citet{2003AJ....126..863G} and \citet{2006ApJ...647L.159Y}.  A solid line
intersecting LFAM~26 is drawn to indicate the $\sim$100$^\circ$ position angle
of the LFAM~26 outflow and is reproduced in the top right inset, which shows
continuum H$_2$ emission from LFAM~26 \citep{2004A&A...427..651D}.  This line
is nearly a beamwidth away from [GSWC2003]~2 (designated G~2 in the figure), an
H$_2$ emission feature detected by \citet{2003AJ....126..863G}.  A solid curve
through the peaks of the Elias~29 outflow lobes indicates the S-shaped symmetry
of the outflow and is weighted by eye to trace the H$_2$ emission from
\citep{2006ApJ...647L.159Y}.  The bottom right inset shows H$_2$ emission and a
10$\sigma$ contour of continuum emission detected in the narrowband H$_2$
filter used for the observations.  The physical scale is shown in the inset
\citep{2006ApJ...647L.159Y}.  The 22$\arcsec$ FWHM beam is shown in the top
left corner. } \label{fig:intensitycontours}

\end{figure}

Figure~\ref{fig:intensitycontours} shows the integrated intensity map of the
blue-shifted (solid contours, $v_{LSR} = -4$ to 1 km~s$^{-1}$) and red-shifted
(dashed contours, $v_{LSR} = 7$ to 12 km~s$^{-1}$) outflow emission.  The
velocity ranges have been chosen to avoid the absorption features from the
intervening clouds detected by \citet{2002ApJ...570..708B}.  Crosses show the
location of Elias~29 and LFAM~26, plus signs indicate the position of infrared
sources \citep{1983ApJ274..698W}, squares represent starless cores
\citep{1998A&A...336..150M}, and the diamonds show H$_2$ emission knots
observed by \citet[feature G2][]{2003AJ....126..863G} and by \citet[features
1-3b][]{2006ApJ...647L.159Y}.  Elias~29 is associated with emission towards the
north (blue-shifted) and south (red-shifted), consistent with results from
previous outflow surveys \citep{1996A&A311..858B,1997ApJ489L..63S} that have
suggested a bipolar outflow with a mostly pole-on inclination
\citep{1995ApJ...438..250H}.  

The large scale outflow morphology strongly resembles the S-shaped symmetry
observed in H$_2$ emission near the driving source \citep{2006ApJ...647L.159Y},
with the blue lobe bending to the northeast and the red lobe bending to the
southwest.  The solid curve through the peaks of the Elias~29 blue and red
lobes represents our best estimate of the outflow axis and is weighted by eye
towards the five H$_2$ emission knots labelled as features 1, 2a, 2b, 3a, and
3b \citep{2006ApJ...647L.159Y} in the main panel of
Figure~\ref{fig:intensitycontours}.  The bottom right inset is reproduced from
\citet{2006ApJ...647L.159Y} and shows the structure of the region as seen in
H$_2$ emission.   Regions $3a$ and $3b$ appear to trace the peak of the
red-shifted outflow emission.  \citet{2002A&A...395..863C} mapped CO~($J=6-5$)
line emission over a $35 \arcsec \times 35 \arcsec$ field centered on Elias~29
with an effective beamsize of $12 \arcsec$.  Contour maps of their observed
wing emission show evidence in support of an east-west outflow axis, consistent
with the S-shaped symmetry found by \citet{2006ApJ...647L.159Y} and seen in the
central regions of our map.  Higher spatial resolution observations will be
necessary to explore the detailed morphology of the central region of outflow
emission from Elias~29 and provide a more precise comparison with the known
H$_2$ regions.

In addition to the well-studied outflow from Elias~29, we report a new outflow
whose source appears to be LFAM~26, a Class~I protostar
\citep{1994ApJ...434..614G,2001A&A...372..173B} located $\sim$1$\arcmin$
northwest of Elias~29.  The morphology of this outflow---with lobes to the east
and west both showing significant red-shifted emission---suggests that its axis
lies near the plane of the sky, as shown by the solid line in
Figure~\ref{fig:intensitycontours}. The relatively low level of blue-shifted
emission from this outflow is most likely due to the presence of the foreground
absorbing clouds at $v_{LSR} = 2.7$~km$\:$s$^{-1}$ and $v_{LSR} =
3.8$~km$\:$s$^{-1}$ \citep{2002ApJ...570..708B}. In particular, the $v_{LSR} =
2.7$~km$\:$s$^{-1}$ cloud is roughly spatially coincident with the eastern lobe
of the LFAM~26 outflow.  H$_2$ emission detected within the contours of the
integrated emission intensity to the east \citep{2004AAS...20513604Y} lends
further support to the notion of LFAM~26 as the driving source.  The outflow
axis as drawn lies nearly a full beamwidth away from [GSWC2003]~2, a knot of
H$_2$ emission detected by \citet{2003AJ....126..863G}.  These authors
attributed the H$_2$ knot to the Elias~29 outflow, but in our CO~($J=3-2$) wing
map, the western lobe of the LFAM~26 outflow appears to lie between LFAM~26 and
[GSWC2003]~2, suggesting that LFAM~26 is a more likely candidate than Elias~29
to be the powering source of the H$_2$ knot.  

It must be kept in mind that this is a particularly dense region of $\rho$-Oph
which makes the association of outflows with their driving sources particularly
problematic.  For this reason, additional evidence linking LFAM~26 and the
observed outflows is desirable.  This evidence comes in the form of nebulous
emission seen in deep NIR imaging of LFAM~26 that has a distinct hourglass
shape---shown in the top right inset of Figure~\ref{fig:intensitycontours}
\citep[taken from Fig.~4 in][]{2004A&A...427..651D}---with the nebulosity
extending a few arcseconds away on either side of the central source.  The
orientation of the hourglass roughly coincides with the $\sim$100$^\circ$
outflow position angle we estimated above, serving as further evidence that
LFAM~26 is the powering source of the observed outflow.  The hourglass shape
around LFAM~26 is consistent with the presence of an edge-on accretion disk.
Whether or not LFAM~26 qualifies as a VeLLO must await detailed modeling of its
IR SED, a task which is beyond the scope of this letter.  

It should be noted that the integrated emission intensity peaks do not appear
to lie directly on the outflow axis as shown in
Figure~\ref{fig:intensitycontours}.  Due to the masking effect of foreground
clouds, the redshifted emission components alone may not accurately portray the
east-west spatial centroids of the outflow.  Nevertheless, let us consider the
possibility of alternative driving source(s).  First, the object closest to
LFAM~26 is E-MM3, located $\sim$15$\arcsec$ east and $\sim$30$\arcsec$ south of
LFAM~26.  This object was first labelled as a starless core
\citep{1998A&A...336..150M}, but subsequent VLT H and Ks observations showed a
bipolar reflection nebulosity surrounding it in much the same manner as LFAM~26
\citep{2000A&A...364L..13B}.  However, the orientation of this reflection
nebula is closer to north/south than east/west and is therefore an unlikely
candidate for the driving source.  Another possibility is that each of the
outflows is driven by one of the other embedded objects either in this region
(e.g. E-MM4 or E-MM5) or beyond.  Given the highly obscured, complex structure
of the region, this scenario is plausible.  Finally, it is conceivable that
Elias~29 is a binary system, with the other object producing the observed
outflow.  However, numerous searches using a wide variety of techniques from
lunar occultations \citep{1995ApJ...443..625S} to deep NIR imaging
\citep{2004A&A...427..651D} have found no evidence for a companion.  

\subsection{Outflow Energetics}\label{sec:energetics}

Physical parameters of each outflow are summarized in Table~\ref{tab:results}.
From the integrated intensity map, we estimate the maximum radius ($R_{max}$)
and area ($A$) of each outflow lobe centered on the position of peak flux.
Here the characteristic velocity of each outflow, $v_{char}$, is the maximum
outflow velocity detected in the associated $^{12}$CO~($J=3-2$) spectra, $\sim
8 \: {\rm km \: s}^{-1}$.  The dynamical timescale is $t_d = R_{max}/v_{char}$.
We assume the wings of the $^{12}$CO~($J=3-2$) emission are optically thin, so
the values in Table~\ref{tab:results} represent lower limits to the actual
values.  As mentioned in Section~\ref{sec:morph}, however, there is evidence of
strong absorption of the blue-shifted outflow emission from LFAM~26 due to the
presence of foreground absorption clouds.  Furthermore, the measured
radial velocity underestimates the characteristic velocity by a factor of
$1/sin(i)$, where $i$ is the inclination angle.  If the outflow axis of LFAM~26
is near the plane of the sky, then $i$ approaches 0$^\circ$ and the correction
factor can increase significantly.

From the peak main beam temperature in the $^{12}$CO~($J=3-2$) spectrum, we
estimate $T_{ex} = 25\:$K.  While $T_{ex}$ is not constant across the map, the
variations are small and do not significantly affect the results.  For
instance, if we assume a $T_{ex}$ that is smaller by a factor of 2, we find
average column densities that are roughly a factor of 2 larger.  The total CO
column density, $N_T$, is calculated from \citet{2005ApJ625..194K} under the
assumption of LTE.  Assuming dark cloud abundances of [H$_2$]/[CO]$=1 \times
10^4$, we estimate the mass of each outflow lobe to be $M_{OF} =
([$H$_2$/[CO])$N_T$$A \mu_g m_{{\rm H}_2}$, where $\mu_g$ is the mean atomic
weight of the gas, 1.36 \citep{1997ApJ489L..63S}.  The mass outflow rate is
$\dot{M}_{OF} = M_{OF} / t_d$, the outflow momentum is $P_{OF}=M_{OF}v_{char}$
and the outflow force is $F_{OF} = p_{OF}/t_d$.  The mechanical luminosity is
estimated as $L_m = \frac{1}{2}M_{OF}v_{OF}^2 / t_d$.  The outflow to infall
conversion efficiency is $\epsilon = \dot{M}_{OF} / \dot{M}_{accr}$.  

\citet{1997ApJ489L..63S} and \citet{1996A&A311..858B} have obtained integrated
intensity $^{12}$CO~($J=2-1$) emission maps centered on Elias~29.  The
$^{12}$CO~($J= 3-2$) map shown here represents a significant enlargement of the
surveyed area.  Inside the mapped region, we find a larger radius for the blue
lobe of Elias~29---0.058~pc compared to 0.024~pc in \citet{1997ApJ489L..63S}.
We also find an average column density over the Elias~29 outflow that is a
factor of $\sim$$2-3$ lower than observed in $^{12}$CO~($J=2-1$).  The
difference in column density estimates may be due to the $^{12}$CO~($3-2$) wing
emission being optically thick. Our optically thin assumption would then
underestimate the true column density by a factor of $\tau/(1-e^{-\tau})$.  Due
to the extended size of the outflow region, the $^{12}$CO~($J=3-2$) derived
outflow mass is comparable to that of \citet{1997ApJ489L..63S}.  The outflow
energetics are summarized in Table~\ref{tab:results}. The low values of the
energetic parameters associated with Elias 29 suggest it is near the end of its
outflow/accretion phase, consistent with the findings of
\citet{2002A&A...395..863C}.

\section{Summary}\label{sec:summary}

We present a 25 square arcminute map of $^{12}$CO~($J=3-2$) outflow emission
centered on Elias~29.  In the map, we see evidence suggesting the Elias~29
outflow is bipolar and has a curved flow that traces the same structure seen in
H$_2$ line emission.  We report the first detection of a molecular outflow from
LFAM~26, an object whose low luminosity makes it a candidate VeLLO.  While the
high density of protostellar objects in this region produces a complicated
outflow structure that makes associating outflows with their driving sources
somewhat problematic, we conclude LFAM~26 is the most likely driving source.
Both the molecular outflow and the NIR scattered light cavity morphology
suggests the associated outflow axis lies near the plane of the sky, in which
case we are viewing the protostar through an edge-on accretion disk.  We
calculate lower limits on the mechanical luminosity of the outflows from
Elias~29 and LFAM~26 to be 6.4 and 10.3~$ \times 10^{-3} L_\sun$, respectively.
The low value of $L_m$ for Elias 29 suggests this object is near the end of its
outflow/accretion phase of evolution.  

We wish to acknowledge the comments of the referee which were very helpful
in refining the paper.

\clearpage

\begin{deluxetable}{lcc}
\tablecolumns{3}
\tablewidth{260pt}
\tablecaption{Parameters of Detected Sources}
\tablehead{
\colhead{Parameter} & \colhead{Elias~29} & \colhead{LFAM~26}
}
\startdata
$R_{max}$ (pc) & 0.058 & 0.064 \\
$t_{d}$ ($10^3\:$yr) & 7.1 & 7.8 \\
$<$$N_{CO}$$>$ ($10^{15}\:$cm$^{-2}$) & 3.6 & 3.3 \\
$A_{OF}$ (sq. arcsec) & 5100 & 6600 \\
$M_{OF}$ ($10^{-3} M_\sun$) & 8.7 & 15.2 \\
$\dot{M}_{OF}$ ($10^{-6} M_\sun$~yr$^{-1}$) & 1.2 & 1.9 \\
$\epsilon$ ($\dot{M}_{OF}/\dot{M}_{accr}$) & 0.15 & 0.25 \\
$P_{OF}$ ($10^{-3} \: M_\sun \:$km$\:$s$^{-1}$) & 69 & 121 \\
$F_{OF}$ ($10^{-6} \: M_\sun \:$km$\:$s$^{-1}$ ) & 9.8 & 15.6 \\
$L_{m}$ ($10^{-3} \: L_\sun$) & 6.4 & 10.3 \\
\enddata
\label{tab:results}
\end{deluxetable}

\clearpage

\clearpage

\end{document}